\documentclass[twocolumn,prb,showpacs,aps,amsmath,superscriptaddress]{revtex4-1}
\usepackage{graphicx}
\usepackage{epsfig,euscript}
\usepackage{lineno}
\usepackage{color}

\usepackage[utf8]{inputenc}
\usepackage[english]{babel}
\usepackage{amsmath}
\usepackage{amsfonts}
\usepackage{amssymb}
\usepackage{fontenc}

\begin{document}
\title{Resonant inelastic x-ray scattering probes the electron-phonon coupling in the spin-liquid $\kappa$-(BEDT-TTF)$_2$Cu$_2$(CN)$_3$}

\author{V. Ilakovac}
\affiliation{Sorbonne Universit\'es, UPMC, Univ. Paris 6, CNRS UMR 7614, Laboratoire de Chimie Physique -- Mati\`ere et Rayonnement, Paris, France}
\affiliation{Universit\'e de Cergy-Pontoise, F-95031 Cergy-Pontoise, France}
\email{vita.ilakovac-casses@upmc.fr}

\author{S. Carniato}
\affiliation{Sorbonne Universit\'es, UPMC, Univ. Paris 6, CNRS UMR 7614, Laboratoire de Chimie Physique -- Mati\`ere et Rayonnement, Paris, France}

\author{P. Foury-Leylekian}
\affiliation{Laboratoire de Physique de Solides, CNRS UMR 8502, Univ. Paris Sud, Universit\'e Paris Saclay, Orsay, France}

\author{S.  Tomi\'c}
\affiliation{Institut za fiziku, P.O.Box 304, HR-10001 Zagreb, Croatia}

\author{J.-P. Pouget}
\affiliation{Laboratoire de Physique de Solides, CNRS UMR 8502, Univ. Paris Sud, Universit\'e Paris Saclay, Orsay, France}

\author{P. Lazi\'c}
\affiliation{Ruder Bo\v{s}kovi\'c Institute, Bijeni\v{c}ka cesta 54, HR-10000 Zagreb, Croatia}

\author{Y. Joly}
\affiliation{Institut N\'eel, 25 Avenue des Martyrs, Grenoble, France}

\author{K. Miyagawa}
\affiliation{Department of Applied Physics, University of Tokyo, Tokyo 113-8656, Japan}

\author{K. Kanoda}
\affiliation{Department of Applied Physics, University of Tokyo, Tokyo 113-8656, Japan}

\author{A. Nicolaou}
\affiliation{Synchrotron SOLEIL, L'Orme des Merisiers, Saint-Aubin, B.P. 48, F-91192 Gif-sur-Yvette, France}

\date{\today}

\begin{abstract}
Resonant inelastic x-ray scattering at the N K edge reveals clearly resolved harmonics of the anion plane vibrations in the $\kappa$-(BEDT-TTF)$_2$Cu$_2$(CN)$_3$ spin-liquid insulator. 
Tuning the incoming light energy at the K edge of two distinct N sites permits to excite different sets of phonon modes. 
Cyanide CN stretching mode is selected at the edge of the ordered N sites which are more strongly connected to the BEDT-TTF molecules, while positionally disordered N sites show multi-mode excitation. 
Combining measurements with calculations on an anion plane cluster permits to estimate the site-dependent electron-phonon coupling of the modes related to nitrogen excitation.
\end{abstract}


\maketitle

\section{\label{sec:INTRO}{Introduction}}

Electron-phonon coupling (EPC) plays a fundamental role in many aspects of 
condensed matter physics. 
It governs the charge mobility and optical properties in metals and semiconductors, 
it drives the Peierls metal-insulator instability in charge density wave compounds, 
it gives rise to the conventional superconductivity \cite{Giustino}, 
while its role in the unconventional superconductivity is under persisting debate. 
Among experimental techniques permitting to measure the EPC strength, like 
infrared (IR), Raman, angle resolved photoemission spectroscopy, 
inelastic neutron scattering, and more recently ultrafast transient response of 
the optical reflectivity \cite{Mansart}, neither of them is at the same time 
element-selective, site-dependent and momentum resolved like 
resonant x-ray scattering (RIXS). 
Last decade improvement of the resolving power of the RIXS spectrometers, 
going up to E/$\Delta$E of 2.5 10$^4$,\cite{Chaix} 
promotes it to an excellent technique for the direct measurement of 
the EPC strength with all these advantages.

\begin{figure}[t]
\begin{center}
\includegraphics[width=8.6cm,angle=0]{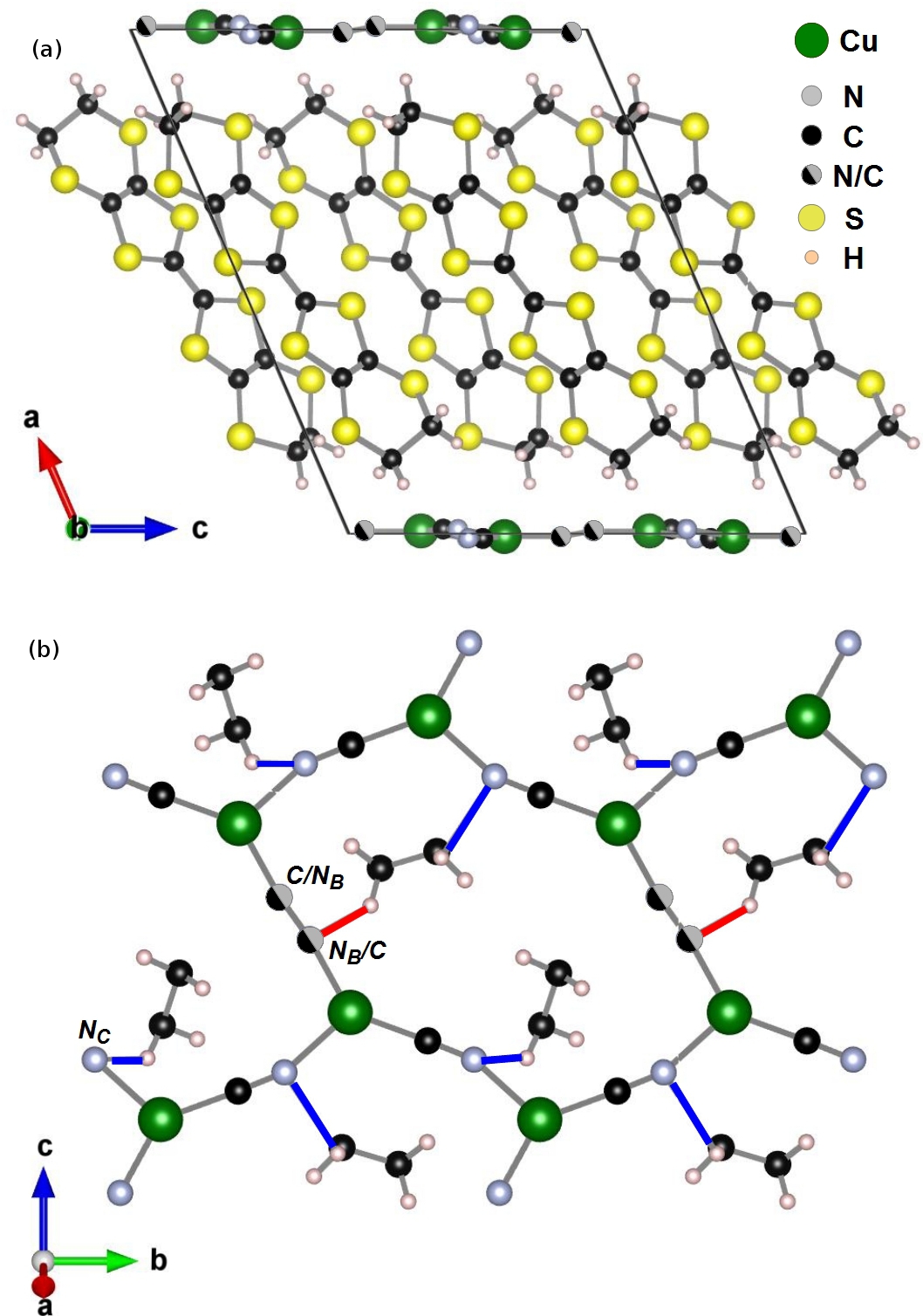}
\caption{\label{fig_1} (a) k-ET-Cu structure  seen along the $b$ axis. (b) Cu$_2$(CN)$_3$ anion plane with Cu-CN$_C$-Cu polymeric chains and bridging, orientationally disordered, CN$_B$ groups. Blue lines indicate HN$_C$ bonds (2.72-2.79 \AA) and red lines HN$_B$ bonds (2.80 \AA) connecting to ethylene endgroups of ET-molecules just below.}
\end{center}
\end{figure}

Understanding how the lattice dynamics couples to charge and spin degrees of freedom 
in $\kappa$-(BEDT-TTF)$_2$Cu$_2$(CN)$_3$ (shortly k-ET-Cu) is of primary importance 
for its spin-liquid  \cite{Shimizu} and pressure-induced superconducting properties 
\cite{Geiser, Komatsu, Kurosaki}. 
This charge-transfer salt is composed of two alternating building blocs, as shown in 
Fig.~\ref{fig_1}(a). 
One is the donor layer of triangular constellations of dimers of ET (BEDT-TTF, 
bisethylenedithio-tetrathiafulvalene) molecules. 
The other is the anion plane, consisting of triangularly coordinated copper(I) ions 
linked by cyanide (CN) groups \cite{Geiser}. 
The two are connected via C-H-N hydrogen bond. 
Each dimer donates approximately one electron to the anion plane, creating a 
triangular lattice of holes with a spin-1/2. 
The absence of spin ordering despite an exchange coupling of J $\approx$ 250~K was 
explained by the total frustration of spins on this triangular lattice \cite{Shimizu}. 
But more recent ab-initio calculations pointed to slightly anisotropic transfer 
integrals between dimers \cite{Kandpal, Jeschke} which rose the question of the 
origin of the spin-liquid state. 
A possible explanation lies in an interaction between spins and charge dynamics 
\cite{Hotta, Ishihara}, and spinon-phonon interaction, as pointed by the ultrasonic 
wave measurements \cite{Poirier}. 
Finally, the actual debate on the role of phonons in the Cooper pairing applies 
perfectly to its pressure induced unconventional superconductivity \cite{Kagawa}.

There are many evidences of the strong dynamical and/or statical disorder in k-ET-Cu. 
An inherent disorder is present already in the conventional P2$_1$/c structure as 
one third of the anion plane CN groups lies on an inversion point and is thus 
orientationally disordered (see Fig.~\ref{fig_1}(b)).  
Moreover, ET-ethylene endgroups can take different conformations relative to the 
rest of the molecule, similarly to other (ET)$_2$X compounds \cite{Alemany2012, Alemany2015}. 
Ethylene hydrogens form stronger H-bonding with nitrogens in ordered polymeric 
chains (N$_C$), compared to disordered bridging sites (N$_B$). 
A subtle interaction between electrons and lattice vibrations is observed in conducting and dielectric properties. 
DC conductivity shows an insulating behavior, described by nearest-neighbor hopping at ambient temperature, variable-range hopping below $\approx$ 130 K, while below 50 K, Hall measurements indicate a complete freezing of charge carriers \cite{Pinteric, Culo}. 
Although there are no electric dipoles associated with the ET dimers \cite{Sedlmeier}, the fingerprints of relaxor ferroelectricity have been established by dielectric spectroscopy below 60 K \cite{Pinteric, Abdel-Jawad}. 

In this work, we focus on the anion plane dynamics as this part of the system controls the donor 
packing and is an intrinsic source of disorder. 
We show that N K edge RIXS permits to switch-on two different dynamics related to nonequivalent 
nitrogen sites which are differently coupled to the ET molecular layer. 
One of these sites shows an essentially mono-mode excitation with five clearly resolved harmonics 
which show up in RIXS spectra. 
Despite the complexity of the system, our RIXS calculations, performed on an anion-plane cluster 
and including vibrational progression, describe well the experimental spectra 
and permit to determine the electron-phonon coupling of the selected phonon modes. 

\section{\label{sec:EXP_CALC}{Experimental and calculation details}}

High quality single crystal sample of k-ET-Cu was grown 
by the  electro-crystallization route \cite{Geiser, Komatsu}. 
N K edge ($\approx$ 400 eV) RIXS measurements were acquired 
at the SEXTANTS beamline (SOLEIL) \cite{Sacchi, Chiuzbaian} 
with an overall energy resolution of 115 meV.  
Near Edge X-ray Absorption Spectroscopy (NEXAFS) measurements were 
performed in the total electron yield mode with the beamline 
resolution set to 80 meV. 
Data were recorded in grazing incidence geometry (inset of Fig.~\ref{fig_3}), 
with the incoming photon polarization 
($\varepsilon$) almost perpendicular to the anion layer 
$\angle$($\varepsilon$, $a^*$) = 20$^\circ$, or $\varepsilon$ $\perp$ $a^*$,
where $a*$ stays for the reciprocal lattice vector perpendicular to the (b,c) plane. 
The orientation of the b and c axis with respect to the scattering plane was not determined.
The sample was measured as introduced in a temperature range from 300 K to 25 K. 
NEXAFS calculations were performed by the Finite Difference Method Near Edge Structure 
(FDMNES) code \cite{Bunau, Guda} using coordinates of the conventional P2$_1$/c structure \cite{Geiser}.  
RIXS calculations were performed on a five-cyanide cluster cut from 
the anion plane, Cu$_2$(CN)$_5$H$_4$. 
Its four peripheral cyanides were ended by hydrogen atoms in order to 
insure the chemical stability, while the central cyanide nitrogen was core-excited. 
Frequencies of the normal modes and the geometry optimization in the 
initial and the intermediate state of the RIXS process were calculated by 
DFT using the GAMESS(US) program \cite{GAMESS}. 
For the quasi-elastic part of the spectra, the calculation of Franck-Condon 
amplitudes was performed by the method using the frequencies of the initial 
and the final state, and taking into account the deformation of the cluster 
from the initial to the intermediate state \cite{DThomas}. 
For the inelastic part of the spectra, the linear-coupling method was applied, 
where the initial and the intermediate state potentials are supposed to be 
harmonic, and of the same frequency.

\section{\label{sec:NEXAFS}{NEXAFS spectra reveal two distinct nitrogen sites} }

\begin{figure}
\begin{center}
\includegraphics [width=7.5cm,angle=0]{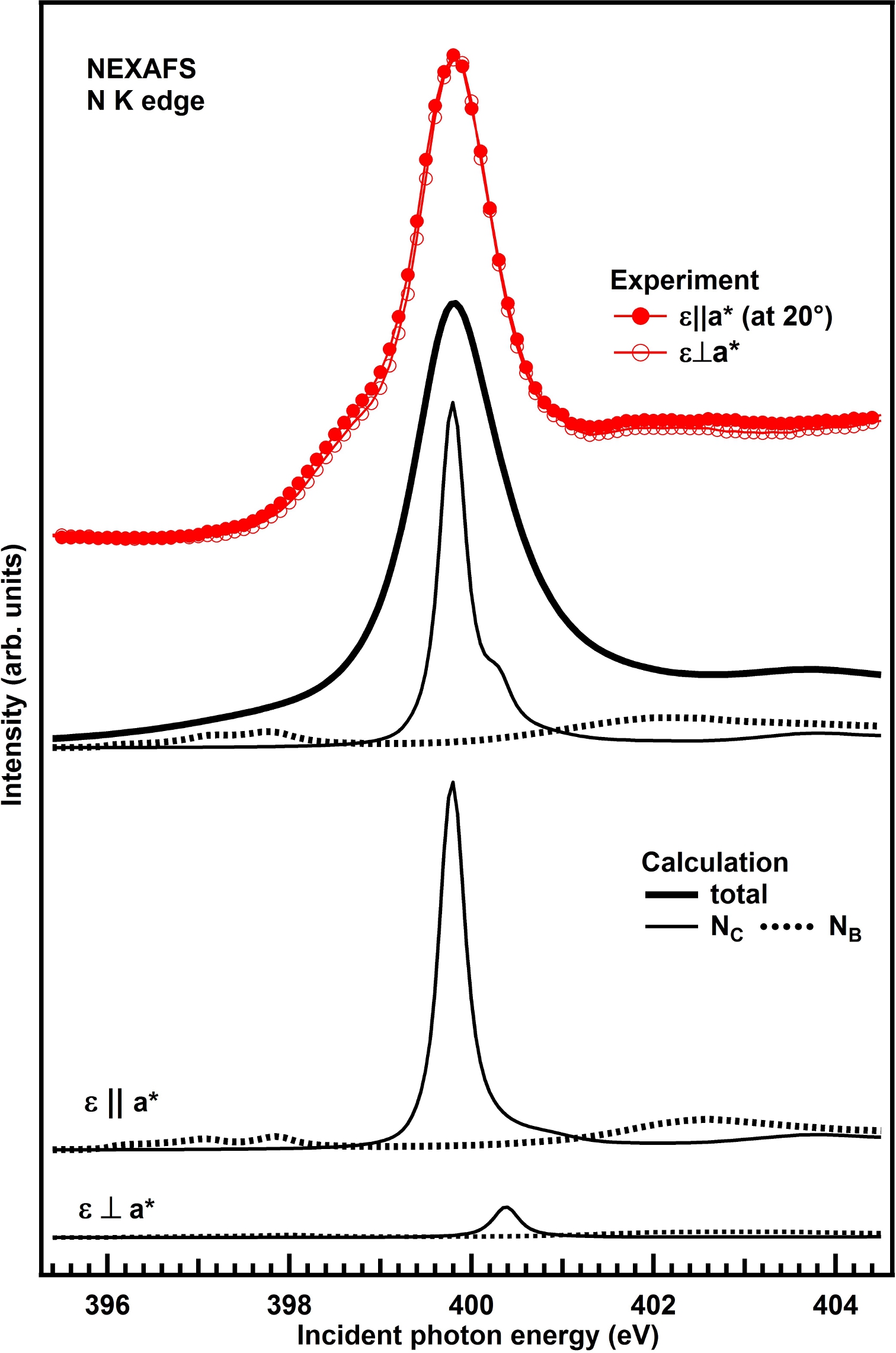}
\caption{\label{fig_2} NEXAFS spectra (red circles) measured for 
$\angle$($\varepsilon$,$a^*$) = 20$^\circ$ and $\varepsilon$ $\perp$ $a*$ 
showing negligible polarization dependence. 
Calculated spectra (thick black line) is performed for 
$\angle$($\varepsilon$,$a^*$) = 20$^\circ$, 
and enlarged for taking into account the dynamics of the system.  
Contributions of N$_C$/N$_B$ sites are shown without enlargement. 
The $\varepsilon$ $\parallel$ $a*$ and $\varepsilon$ $\perp$ $a*$ contributions of 
N$_C$/N$_B$ sites is shown below. } 
\end{center}
\end{figure}

Low temperature N K edge NEXAFS spectrum of the single crystal of k-ET-Cu is shown in 
Fig.~\ref{fig_2}. 
It has a prominent structure at the photon energy of h$\nu$ = 399.8~eV and a low energy shoulder at about h$\nu$ = 398.4~eV.
In order to identify these two features, we performed calculation using the FDMNES code, 
for the two types of N sites (N$_B$, N$_C$) and for two orientations of the incident 
light polarization, parallel and perpendicular to $a^*$. 
The shape of N$_B$ and N$_C$ spectra is very different because the local environment 
of these two nonequivalent sites is not the same. 
For the same reason they also present a relative core level shift, 
higher for N$_C$ compared to N$_B$. 
Indeed, the cation-ET-molecular layer imposes constraints to the anion plane, 
resulting in an elongation of the bridging Cu-CN$_B$-Cu and a compression and bending of the chain Cu-CN$_C$-Cu links (see Tab.\ref{tab1}).
The negative shift of the N$_B$ feature of $\Delta E$ $\approx$ 2.5 eV compared to the main N$_C$ structure is however overestimated compared to the experimental value of $\approx$ 1.5 eV. 
The reason for this discrepancy is the description of the orientational disorder of bridging CN$_B$ cyanide groups in the conventional P2$_1$/c structure \cite{Geiser}. 
In this structure, CN$_B$ groups are centered at inversion points. 
The CN$_B$ distance is determined as the distance between statistical positions of 
C/N$_B$ and N$_B$/C atoms, and is unnaturally increased. 
Moreover, the symmetry imposes equal Cu-C and Cu-N$_B$ distances for a cyanide CN$_B$. 
On the other hand, in the case of chain Cu-CN$_C$-Cu links, the carbon is closer to its copper-neighbor and that the CN$_C$ distance is shorter.

\begin{table}
\begin{center}
\begin{tabular}{lcccccc}
\hline
\hline
   & CN & CuN & CuC & CuCu & $\angle$CuNC & $\angle$CuCN \\
\hline
$crystal$     \\ 
Chain   & 1.130 & 2.026 & 1.872 & 4.933 & 158.9 & 177.3 \\  
Bridge  & 1.188 & 1.895 & 1.895 & 4.962 & 169.4 & 169.4 \\
\hline
$cluster$ &       &       &       &       &       &  \\ 
initial state & 1.156 & 1.928 & 1.931 & 5.013 & 176.2 & 178.8 \\  
excited state & 1.195 & 1.991 & 1.845 & 5.026 & 176.1 & 178.9 \\
\hline
\hline
\end{tabular}
\caption{\label{tab1} Anion layer interatomic distances (in \AA) and angles (in $^\circ$) in the conventional crystal P2$_1$/c structure \cite{Geiser}. They are compared to equivalent distances/angles in the cluster Cu$_2$(CN)$_5$H$_4$ used for RIXS calculation, in its initial and excited state.}
\end{center}
\end{table}

Further, the linear dichro\"{i}sm of the k-ET-Cu NEXAFS is surprisingly negligible, 
i.e. there is almost no polarization dependence when comparing spectra with 
$\angle$($\varepsilon$,$a^*$) = 20$^\circ$ and $\varepsilon$ $\perp$ $a*$ 
($\varepsilon$ $\parallel$ to $(b,c)$ plane), as shown in Fig.~\ref{fig_2}.
Though, NEXAFS spectra of a similar system, planar nitrile molecules, 
show two prominent white lines, corresponding to excitations to $\pi_{\perp}^*$ 
and $\pi_{\parallel}^*$ states, 
signs ${\perp}$ and ${\parallel}$ meaning relative to the molecular plane. 
Their relative shift of about 1 eV is explained by a conjugation-interaction 
inside the linear arrangement of C=C and C$\equiv$N bonds 
\cite{Carniato2004, Carniato2005, Ilakovac2008, Ilakovac2012}. 
Linear C=C-C$\equiv$N geometry and a strictly planar system are therefore two 
conditions for a splitting of $\pi_{\perp}^*$ and $\pi_{\parallel}^*$ 
which are nominally degenerate. 
In FDMNES calculations of k-ET-Cu spectrum, N$_C$ spectra projected to the 
$\pi_{\perp}^*$ and $\pi_{\parallel}^*$ states, ${\perp}$ and ${\parallel}$ 
meaning relative to the anion plane, are 
separated by only 0.5 eV. 
The conjugation effect is here decreased already by the 30$^\circ$ deviation from 
the linear arrangement of the Cu-CN$_C$-Cu bonds. 
The residual disagreement between the experiment and the FDMNES calculations in terms 
of the polarization dependence can be explained by the DFT approach. 
It does not describe correctly the effect of hydrogen bonding between the anion layer 
nitrogen and ethylene endgroup hydrogen atom. 
Note that this effect should be stronger for N$_C$ (H-N$_C$ = 2.72-2.79 \AA) 
compared to N$_B$ (H-N$_B$ = 2.80 \AA). 

{\textit{We stress the fact that even if the match between the experimental and 
the calculated 
NEXAFS spectra is not perfect, comparing them permits us to identify the NEXAFS 
main peak as related 
to the N$_C$ sites and the low energy shoulder to the N$_B$ sites.}}

\section{\label{sec:Quasi_elast_RIXS}{Site-dependent lattice motion fingerprints in the quasi-elastic part of the RIXS spectra}}

\begin{figure}
\begin{center}
\includegraphics[width=8.5cm,angle=0]{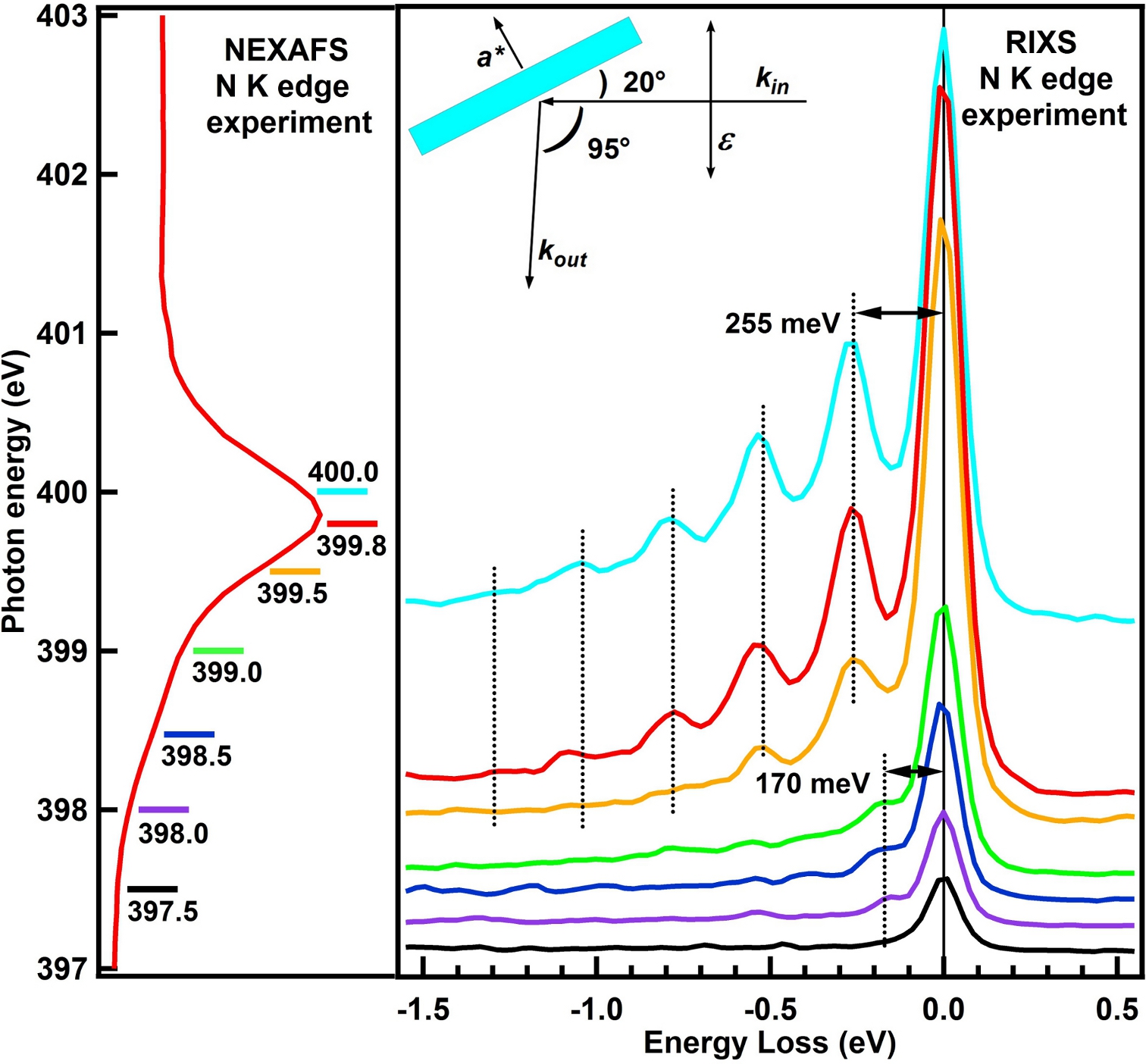}
\caption{\label{fig_3} Low temperature NEXAFS (60 K, left) and RIXS (25 K, right) spectra, 
shifted for clarity. Inset shows the experimental geometry, with the polarization vector 
$\varepsilon$ in the scattering plane.}
\end{center}
\end{figure}

Low temperature N K edge RIXS spectra are shown in Fig.~\ref{fig_3}.  
The experimental geometry is indicated in the inset. 
RIXS spectra were collected at photon energies indicated on the NEXAFS spectrum by the line of 
the same color. 
For incoming photon energies above h$\nu$ = 397.5~eV, the elastic peak 
develops an asymmetric tail, which transforms into a clear vibrational progression, 
concerning directly nitrogen sites of the anion CN groups. 
Recent detailed study of the k-ET-Cu lattice vibrations \cite{Dressel2016} reports that 
there are essentially three distinct families of modes affecting the CN groups. 
Their energy quanta will be designated by $\hbar\omega_n$, with $n$ = 1,2,3. 
They are identified as CN stretching ($\hbar\omega_1$ = 260-265~meV), CN sliding between 
Cu atoms ($\hbar\omega_2$ = 60-65~meV), and low energy modes combining CN bending, 
stretching and twisting ($\hbar\omega_3$ = 8-50~meV), as presented in Tab.~\ref{tab2}. 
The vibrational progression at h$\nu$ = 399.5-400.0~eV shows distinct and perfectly 
resolved harmonics, separated by 255~meV, indicated by equidistant dotted lines. 
They are attributed to CN stretching motion ($\hbar\omega_1$). 
Below the NEXAFS maximum, in the range h$\nu$ = 398-399~eV 
(NEXAFS shoulder), an energy loss of 170~meV is revealed (see Fig.~\ref{fig_3}(right)). 
Its energy does not correspond to any simple CN group movement and points rather 
to a multi-mode excitation. 

\begin{figure}
\begin{center}
\includegraphics[width=8.6cm,angle=0]{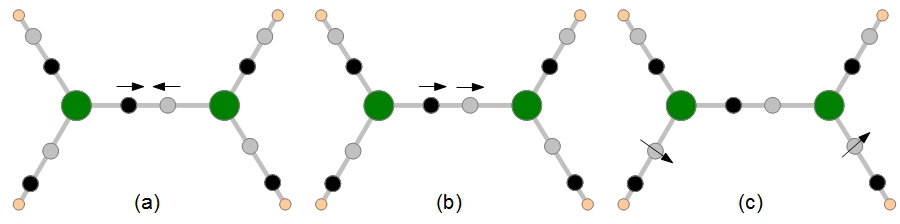}
\caption{\label{fig_4} Cu$_2$(CN)$_5$H$_4$ cluster used for RIXS calculations and its most 
important normal modes: (a) CN stretching $\hbar\omega_1$, (b) CN sliding $\hbar\omega_2$, 
(c) CN bending $\hbar\omega_3$.}
\end{center}
\end{figure}

\begin{figure}
\begin{center}
\includegraphics[width=8.6cm,angle=0]{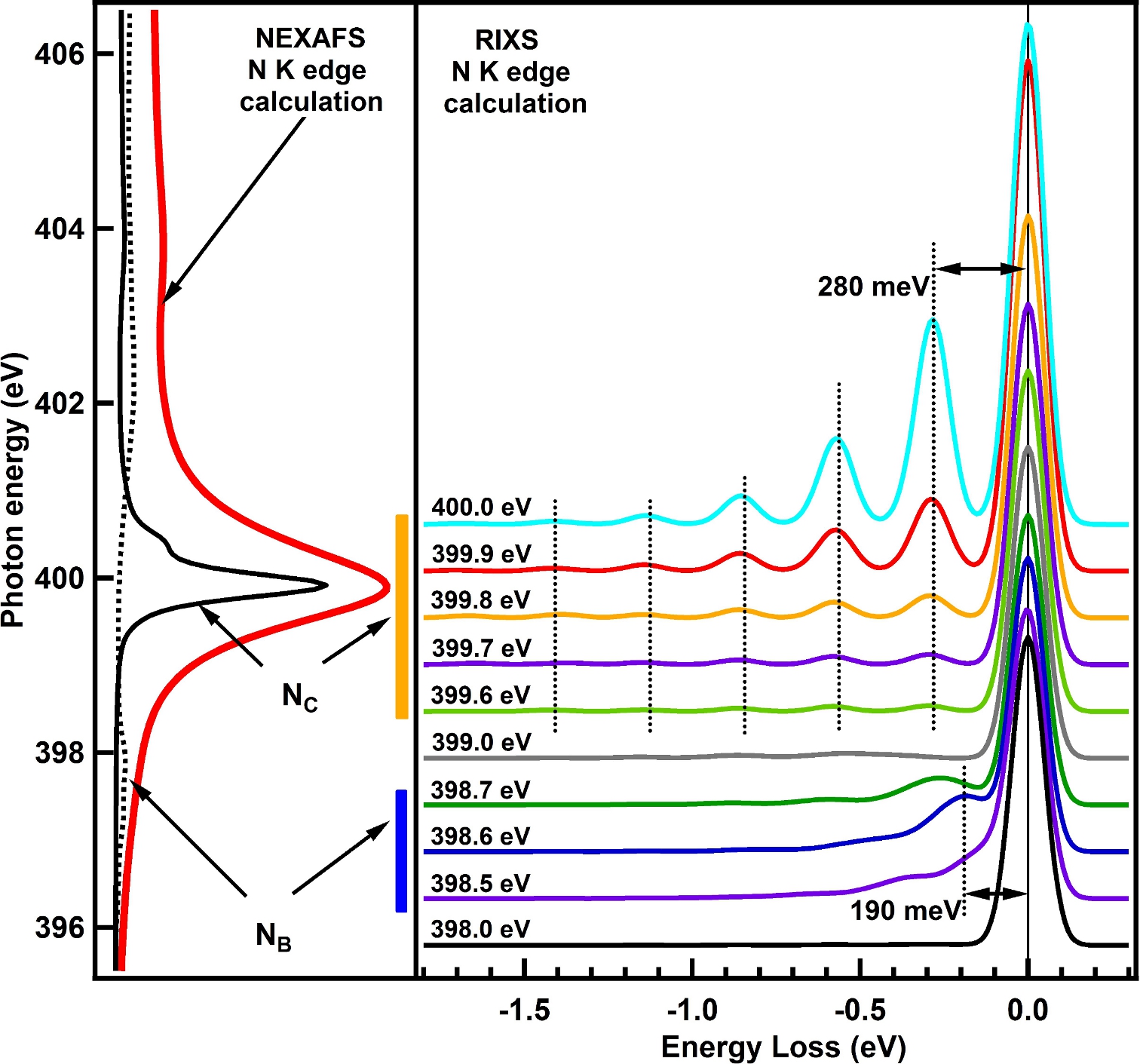}
\caption{\label{fig_5} (left) NEXAFS calculation (red line) and N$_C$ and N$_B$ contributions. 
(right) RIXS calculation on a Cu$_2$(CN)$_5$H$_4$ anion plane cluster.}
\end{center}
\end{figure}

\begin{table}
\begin{center}
\begin{tabular}{lcccc}
\hline
\hline
Mode   & CN stretching & CN sliding  & CN bending \\
$n$  & 1 & 2 & 3\\
\hline
$crystal$ \cite{Dressel2016}  & k-ET-Cu &   &   \\
$\sigma$ &  2100-2140 & 508-515 & 195-225\\
$\hbar\omega$  &  260-265 & 63-64 & 24-28\\
\hline
$cluster$ & Cu$_2$(CN)$_5$H$_4$ & $calculations$  & \\
$\sigma_n$ (6\%)   & 2258 & 510 & 200\\
$\hbar\omega_n$ (6\%)  & 280 & 64 & 25\\
$S_n$ (6\%) & 0.334 & 0.608 & 0.235\\
$g_n$ (9\%) & 230 & 71 & 17 \\
$\lambda_n$ (35\%) & 0.076 & 0.032 & 0.004 \\
\hline
\hline
\end{tabular}
\caption{\label{tab2} Most important vibrational modes involving anion CN groups 
(see Fig.~\ref{fig_4}). Cluster calculation wave numbers ($\sigma$ in cm$^{-1}$) 
and energy quanta ($\hbar\omega$ in meV) are compared to these in the crystal \cite{Dressel2016}. 
Corresponding values of Huang-Rhys parameters ($S$),  gradients in dimensionless 
normal coordinates ($g$ in meV) and dimensionless EPC ($\lambda$) are given with 
their relative errors in parentheses (\%) \cite{errors}.}
\end{center}
\end{table}

In order to reproduce these experimental data we have performed RIXS calculations using 
the Cu$_2$(CN)$_5$H$_4$ cluster depicted in Fig.~\ref{fig_4}. 
It has been chosen as its interatomic distances and normal mode frequencies match well 
reported values in the k-ET-Cu crystal \cite{Geiser, Dressel2016} (see Tab.\ref{tab1} 
and Tab.\ref{tab2}). 
However, it does not include effects of the ET-cation layer. 
The cluster has naturally linear Cu-CN-Cu arrangement characteristic of the C$\equiv$N bond,  
while in the real crystal, stronger H bonding of ethylene endgroups to the N$_C$- 
sites drags nitrogen in polymeric chains and deforms (bends) Cu-CN$_C$-Cu links (see Fig.~\ref{fig_1} (b)). 
For Cu-CN$_B$-Cu bridging sites this deformation is much smaller. 

Considering the grazing-incidence geometry explored in the experiment 
(see inset of Fig.~\ref{fig_3}), we considered that in the intermediate state 
the core electron is accommodated in the $\pi_{\perp}^*$ orbital system 
(perpendicular to the anion plane), while the initial and the final electronic 
states are the same. 
The calculated value of the 1s $\rightarrow$ $\pi_{\perp}^*$  transition of the 
Cu$_2$(CN)$_5$H$_4$ cluster is 398.6.0~eV, considered as the resonance of the N$_B$- sites. 
The resonance of the N$_C$- sites was set to 1.5 eV higher energy, corresponding to the 
energy difference between the two distinct behaviors observed in the RIXS experiment. 
In terms of phonon excitation, the final state differs strongly from the initial situation 
even if only few modes are relevant for the description of the quasi-elastic part of 
the spectra: CN stretching ($\hbar\omega_1$), CN sliding ($\hbar\omega_2$), 
and CN bending ($\hbar\omega_3$), depicted in Fig.~\ref{fig_4}. 

In order to compare experimental data with calculations, calculated NEXAFS 
and RIXS spectra are shown in Fig.~\ref{fig_5} in the same manner as the 
experimental results in Fig.~\ref{fig_3}. 
RIXS calculation shows that the resonant behavior of the vibrational progression 
calculated by 
this simple cluster model agrees well with the experiment.

\begin{figure}
\begin{center}
\includegraphics[width=8.6cm,angle=0]{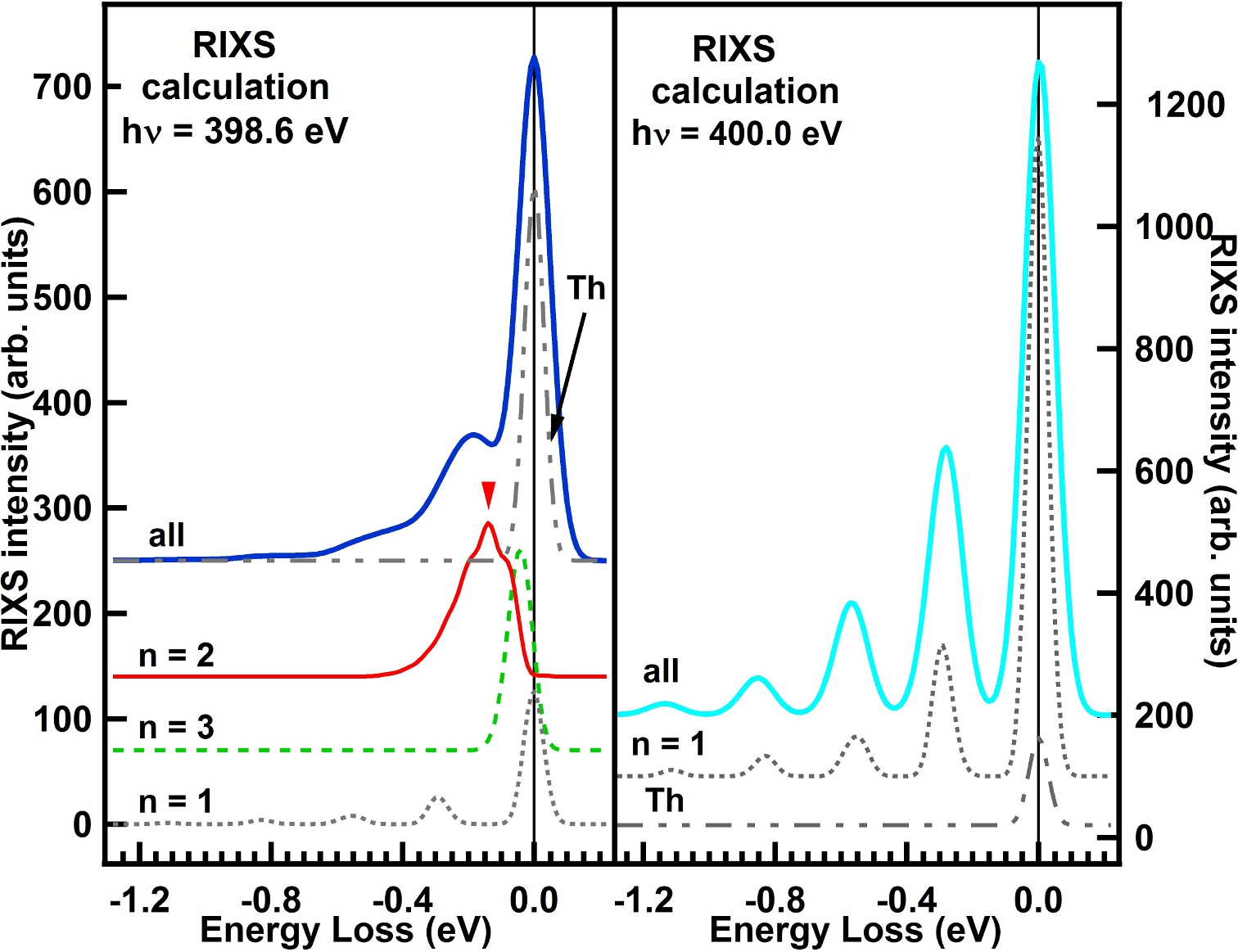}
\caption{\label{fig_6} RIXS spectra calculated for $h\nu$ = 398.6 eV 
(N$_B$ sites, NEXAFS shoulder) and $h\nu$ = 400.0~eV (N$_C$ sites, NEXAFS main peak). 
Each spectrum is decomposed in its mono-mode contributions (n = 1-3) and the 
Thompson scattering (Th). For $h\nu$ = 398.6 eV the second, most excited harmonics 
of the CN sliding  motion (n = 2) is indicated by an arrow. Note that the ``all'' 
spectra include interference effects and experimental resolution broadening.}
\end{center}
\end{figure}

\begin{figure}
\begin{center}
\includegraphics [width=8.cm,angle=0]{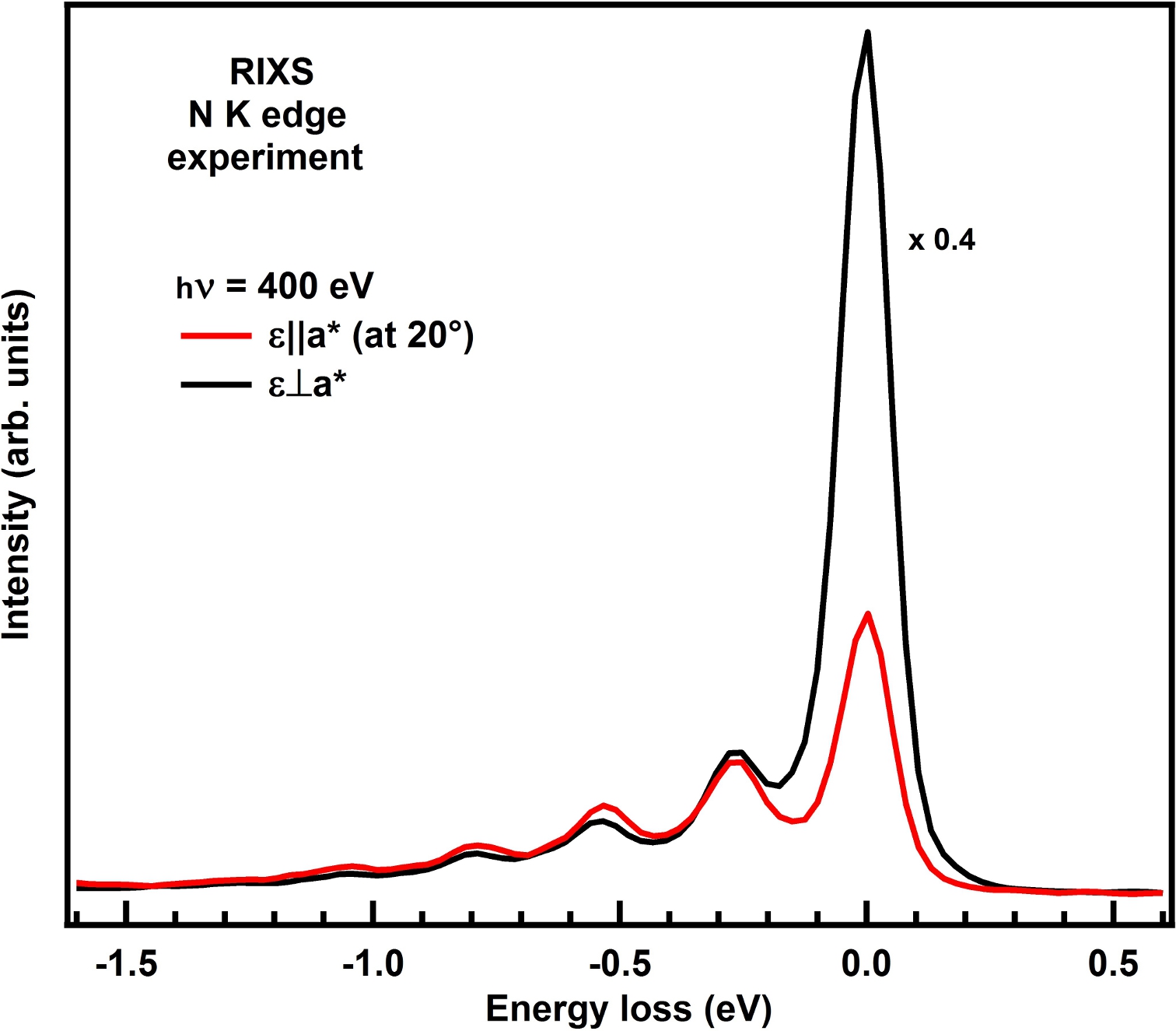}
\caption{\label{fig_7} Experimental RIXS spectra measured at h$\nu$ = 400~eV for two 
orientations of the incident electric field, $\angle$($\varepsilon$,$a^*$) = 20$^\circ$ 
and $\varepsilon$ $\perp$ $a*$. In order to better compare the two vibrational progressions, 
the intensity of the $\varepsilon$ $\perp$ $a*$ spectrum is multiplied by 0.4.}
\end{center}
\end{figure}

When the RIXS calculation is performed for the photon energy $h\nu$ = 398.6 eV, 
corresponding to the resonance of the N$_B$ sites (NEXAFS shoulder), three normal 
modes are strongly excited. 
Fig.~\ref{fig_6}(left) shows the mono-mode contributions of the three modes ($n$ = 1-3), 
elastic peak (Thomson scattering), and how they all interfere in the RIXS spectra (``all''). 
On the other hand, calculation with the photon energy set to $h\nu$ = 400 eV 
(N$_C$ sites, NEXAFS main peak), shown in Fig.~\ref{fig_6}(right) matches the 
experiment when solely CN stretching ($\hbar\omega_1$) is included. 
Clearly, real-crystal constraints in the curved Cu-CN$_C$-Cu conformation prevail 
strong excitation of any other motion related to the CN$_C$ group. 
Strong excitation of any other mode would indeed completely destroy the CN stretching 
vibrational progression with distinct harmonics, similarly to the case of the N$_B$ 
excited sites in the Fig.~\ref{fig_6}(left). 
Moreover, Cu-N and Cu-C distances in the real crystal are already close to these of the N K 
edge excited state cluster. 
Small excited state cluster deformation relative to the real crystal Cu-CN$_C$-Cu 
conformation explains the lack of vibrational modes besides highest energy CN stretching.
Weak excitation of low-energy modes is however not excluded and should contribute 
to the tail of the elastic peak observed in the experiment. 
Finally, changing the polarization to $\varepsilon$ $\perp$ $a^*$ does not change 
the vibrational progression, only the intensity of the elastic peak (see Fig.~\ref{fig_7}). 

{\textit{Observed resonant modification of the vibrational progression in the 
quasi-elastic part of the N K edge RIXS spectra confirms that the NEXAFS main 
peak and the NEXAFS shoulder are related to two 
different nitrogen sites, N$_C$ and N$_B$, respectively. 
With the help of RIXS calculations, we determine that the CN$_C$ groups are restricted 
to almost mono-mode motion, while CN$_B$ groups show up at least three strongly excited modes. }}

\section{\label{sec:Inelastic_RIXS}{Inelastic part of the RIXS spectra - charge transfer excitations}}

\begin{figure}
\begin{center}
\includegraphics [width=8.2cm,angle=0]{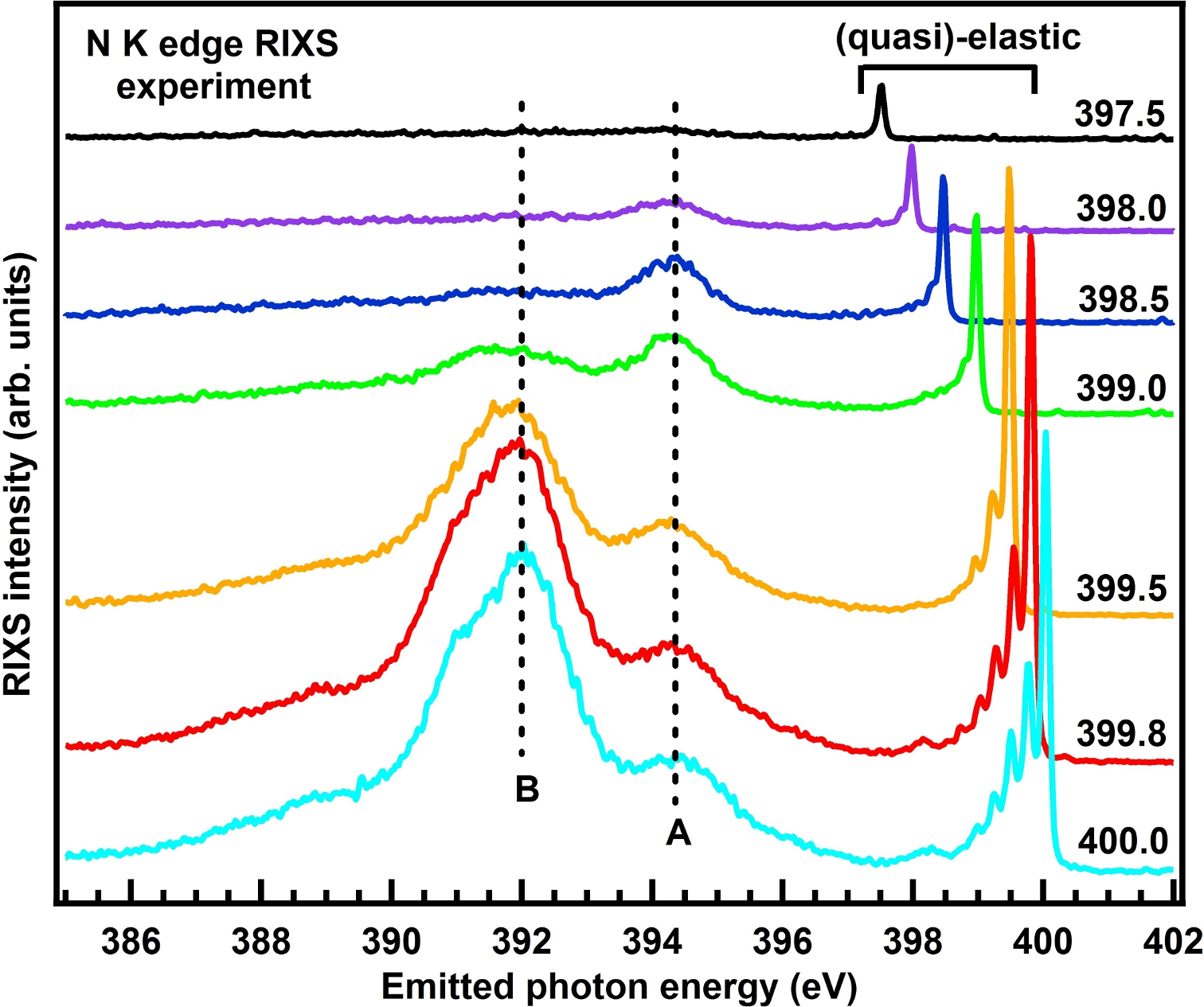}
\caption{\label{fig_8} Emitted-energy scale (h$\nu$') RIXS spectra measured at 25 K. 
Corresponding incident photon energy (h$\nu$) is given at the right side of each spectra. 
Dashed line indicates the emitted energy of the structures A and B position at the 
NEXAFS maximum (h$\nu$ = 399.8 eV).}
\end{center}
\end{figure}

Large energy-range RIXS spectra are shown in Fig.~\ref{fig_8}. 
They are presented in the emitted-energy scale in order to point to the almost 
non-Raman behavior of  inelastic features A and B. 
Feature A appears at slightly lower incident photon energy (h$\nu$) and has almost 
constant intensity in a range of about 2 eV, 
while B resonates at h$\nu$ = 399.8 eV, corresponding to the NEXAFS maximum.
The emitted photon energy of their maximum is 
h$\nu$' $\approx$ 394.5 eV and $\approx$ 392.0 eV, respectively. 

\begin{figure}
\begin{center}
\includegraphics [width=8.cm,angle=0]{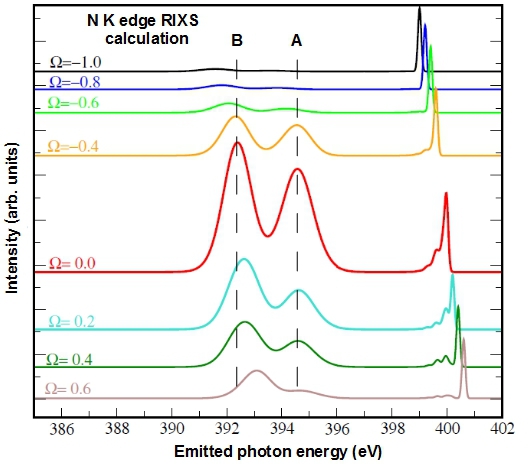}
\caption{\label{fig_9} Calculation of the RIXS spectra including inelastic events. 
The detuning energy $\Omega$ (in eV), relative to the NEXAFS maximum, is indicated for each spectum. 
Dashed line indicates the position of the structures A and B for $\Omega$ = 0.}
\end{center}
\end{figure}

\begin{figure}
\begin{center}
\includegraphics [width=8.5cm,angle=0]{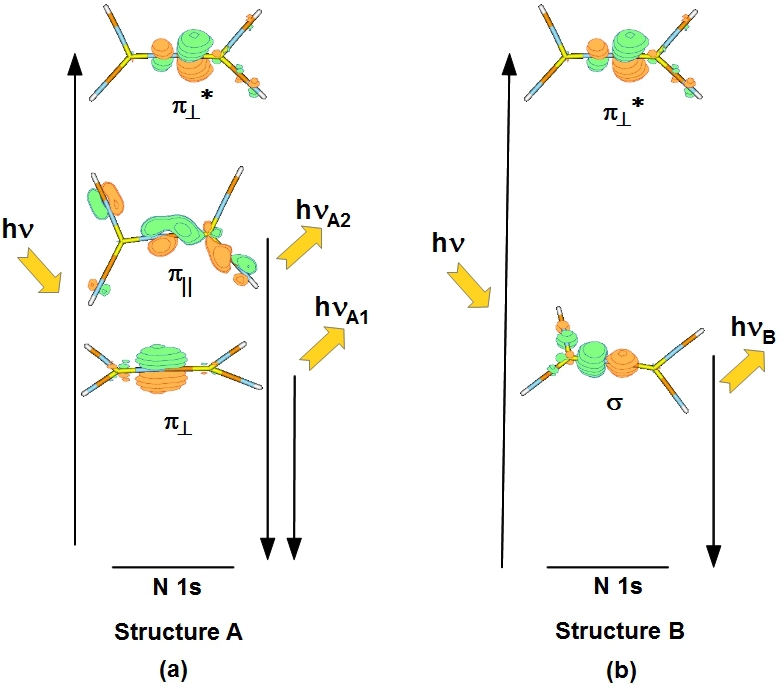}
\caption{\label{fig_10} Inelastic RIXS processes giving rise to the structures A and B, 
with emitted photon energy, $h\nu'$ = $h\nu_{A1}$, $h\nu_{A2}$ and $h\nu_{B}$. 
Lobes of the $\pi_{\perp}$, $\pi_{\parallel}$ and $\sigma$ orbitals 
are presented in the initial state, the $\pi_{\perp}^*$ orbital in the 
excited state.}
\end{center}
\end{figure}

RIXS calculations are performed on the Cu$_2$(CN)$_5$H$_4$ cluster (see Fig.~\ref{fig_9}), 
as for the quasi-elastic part. 
Structure A is identified as the excitation of an N 1s electron 
to the first non-occupied orbital ($\pi_{\perp}^*$) followed by the 
$\pi_{\perp}$, $\pi_{\parallel}$ $\rightarrow$ N 1s  decay, 
giving rise to h$\nu_{A1}$, h$\nu_{A2}$ emission, respectively, 
as schematized in Fig.~\ref{fig_10}(a).  
In terms of initial and final states, it is thus a signature of two 
transitions: 
$\pi_{\perp}$ $\rightarrow$ $\pi_{\perp}^{*}$ and 
$\pi_{\parallel}$ $\rightarrow$ $\pi_{\perp}^{*}$. 
The $\pi_{\perp}$ $\rightarrow$ $\pi_{\perp}^{*}$ transition 
is potentially related to the ET-cation $\rightarrow$ anion charge transfer, 
which is not described in the present anion-cluster calculations. 
The $\pi_{\parallel}$ $\rightarrow$ $\pi_{\perp}^{*}$ transition is a charge transfer from 
an orbital which is delocalized on three neighboring cyanides, to a $\pi_{\perp}^{*}$, which 
is localized mostly on the carbon atom of the central (excited) cyanide. 
The former is possibly related to the effective EPC including ET-molecular layer, 
while the latter is connected with the intra-anion-plane electron-phonon interaction. 
Similarly, the structure B with lower emitted energy (h$\nu_{B}$), 
can be described as the $\sigma$ $\rightarrow$ $\pi_{\perp}^{*}$ transition 
(see Fig.~\ref{fig_10}(b)). 

As the Cu$_2$(CN)$_5$H$_4$ cluster is a system with discrete energy levels, 
features A and B are supposed to show Raman behavior, i.e. to shift in the 
emitted photon energy ($h\nu$') scale, in the same manner as the elastic peak. 
This shift is indeed present in the calculated spectra of Fig.~\ref{fig_9} for incident photon energies ($h\nu$) away from the resonance. 
When the dynamics of the system is included via vibrational-progression calculations, 
their position remains constant 
when approaching the resonance, for h$\nu$ $\approx$ 399.8 eV. 
Similar constant emitted-energy resonant behavior is observed in RIXS spectra of a 
very small system as the HCl molecule \cite{Simon}. 
Thus, even for small systems, strong EPC induces non-Raman behavior in RIXS spectra.

{\textit{In the inelastic part of the N K edge RIXS spectra 
three charge transfer excitations are identified: 
$\pi_{\perp}$ $\rightarrow$ $\pi_{\perp}^{*}$ and 
$\pi_{\parallel}$ $\rightarrow$ $\pi_{\perp}^{*}$ closer to the elastic peak, and 
$\sigma$ $\rightarrow$ $\pi_{\perp}^{*}$ with about 2 eV lower emitted energy. 
Cluster calculation including dynamics shows that their Raman behavior is suppressed at resonance.}}

\section{\label{sec:EPC}{Extracting the electron-phonon coupling from the RIXS spectra}}

Theoretical work of Ament et al. \cite{Ament} indicates how the EPC can be estimated  
from the envelope of the RIXS spectra vibrational progression.
In the case of localized Einstein phonon modes, a simple expression is derived under 
ultra-short core-hole lifetime approximation ($\Gamma$ $\gg$~$\hbar\omega$). 
It relates EPC to the relative intensity of the first and the second harmonics, 
the core-hole lifetime broadening ($\Gamma$), and the phonon quantum ($\hbar\omega$). 
In this way, EPC of selected phonon modes was extracted from the Ti L edge RIXS 
spectra of titanates \cite{Moser, Fatale}. 
For the edge we are dealing with, this approximation is not valid, 
as both values of $\Gamma$ reported in the literature, 
93~meV \cite{Campbell} and 132~meV \cite{Campbell,Neeb}, are lower than 
$\hbar\omega_1$ = 250~meV. 

\begin{figure}
\begin{center}
\includegraphics [width=5.5cm,angle=0]{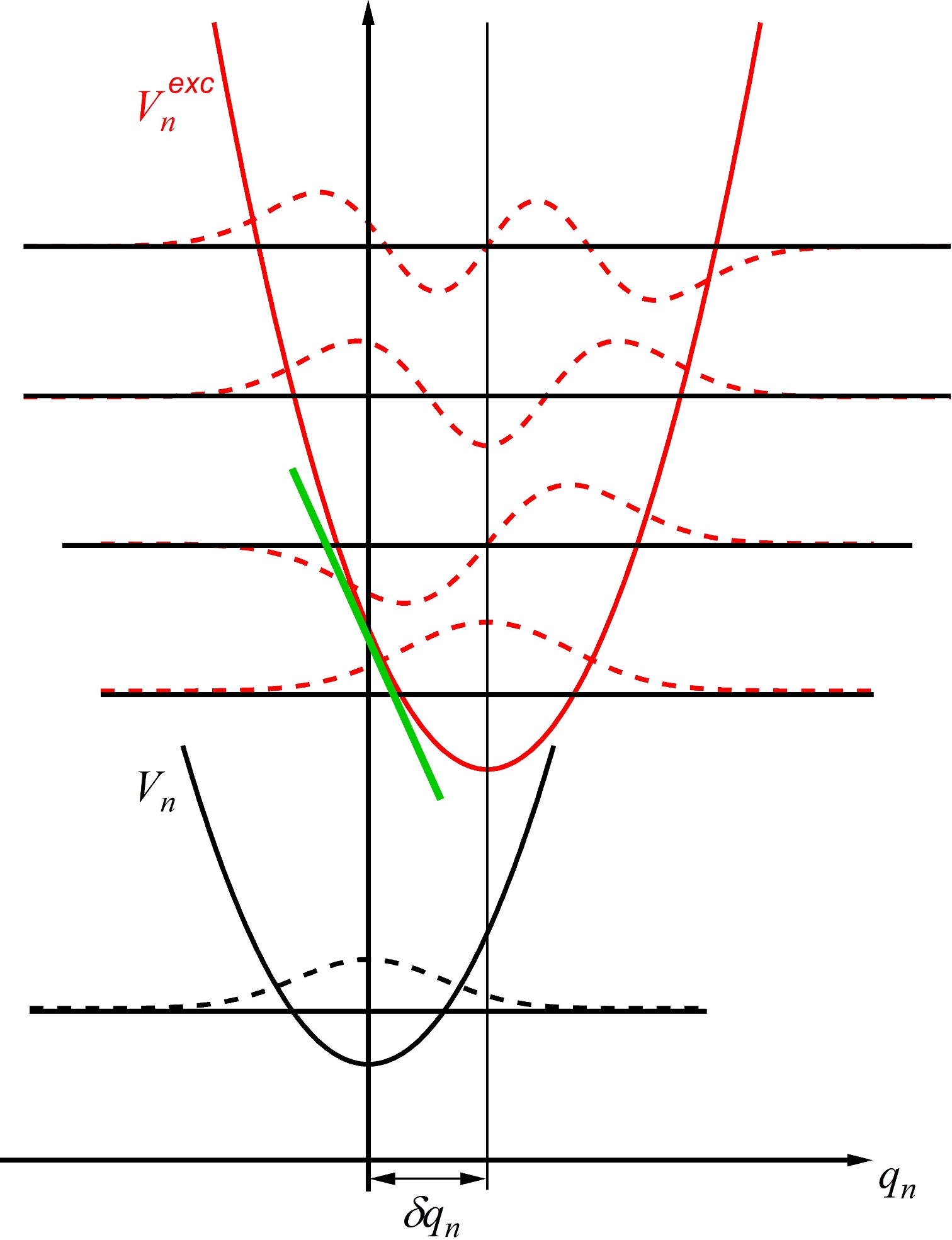}
\caption{\label{fig_11} Initial state (black) and excited state (red line) potential 
and wave functions. 
Gradient of the excited state potential (V$_n^{exc}$) at q$_n$ = 0 is indicated by a green line.}
\end{center}
\end{figure}

Alternatively, the EPC of a selected mode $n$ can be derived from our Cu$_2$(CN)$_5$H$_4$ 
cluster calculations. 
The demonstration will be performed using the linear-coupling model, 
where the potentials in the initial 
and the excited state, are supposed to be harmonic, of the same frequency, $\omega_n$. 
The excited state is here a state with a core-hole and a supplementary electron in the 
valence band.  
Fig.~\ref{fig_11} shows the two electronic-state potentials with corresponding vibrational 
wave functions. 
The phonon coupling to this electronic excitation is evaluated by the overlap of the initial 
state and the excited state wave functions. 
The excited state is shifted relatively to the initial state for a value of $\delta q_n$ in 
normal coordinates.
For small $\delta q_n$, the overlap of the initial and the excited state fundamental prevails, 
and the vibronic excitation is small. 
But, as $\delta q_n$ increases, higher harmonics of the excited state have stronger overlap 
with the initial state wave function, leading to a stronger vibronic coupling. 
The Huang-Rhys (HR)  parameter \cite{HR} evaluates overlaps of wave functions of the two 
harmonic potentials. 
It is calculated from the normal mode coordinate shift $\delta q_n$ of the two potentials 
and their frequency:

\begin{equation}
S_n = \frac{\omega_n}{2 \hbar} \delta q_n^2
\end{equation}

For the quasi-elastic part of the spectra, $\delta q_n$ is calculated by taking into 
account the deformation of the cluster from the initial to the intermediate state \cite{DThomas}.
Nevertheless, deriving of the EPC parameter is more evident when $\delta q_n$ is calculated 
from the gradient of the excited state potential at $q_n$ =  0 
(green line in Fig.~\ref{fig_11}) as was done for the inelastic part of 
the spectra:

\begin{equation}
\delta q_n = \frac{1}{\omega_n^2} {{\left( \frac{\partial V_n^{exc}}{\partial q_n }\right)}_{q_n=0}}
\end{equation}

Combining (1) and (2), $S_n$ can be expressed as:

\begin{equation}
S_n = \frac{1}{2 \hbar \omega_n^3} {{\left( \frac{\partial V_n^{exc}}{\partial q_n}\right)}^2_{q_n=0}}
\end{equation}

Using the transformation to dimensionless normal coordinates $q_n \rightarrow {\sqrt{\hbar / \omega_n}} Q_n $, $S_n$ simplifies to:

\begin{equation}
S_n = {\frac{1}{2}} {\frac{1}{\left(\hbar \omega_n \right)^2}}
{{\left( \frac{\partial V_n^{exc}}{\partial Q_n}\right)}^2_{Q_n=0}}
\end{equation}

This permits to express the gradient of the excited potential in dimensionless normal coordinates 
directly in terms of the HR parameter:

\begin{equation}
g_n = {{\left( \frac{\partial V_n^{exc}}{\partial Q_n}\right)}_{Q_n=0}} = \hbar \omega_n \sqrt{2 S_n}
\end{equation}

In adiabatic EPC systems, where electron dynamics is much faster than 
phonon dynamics ($\omega_{el}$ $\gg$ $\omega_n$), 
the dimensionless EPC constant $\lambda_n$ is defined, through BCS-type theory, as:

\begin{equation}
\lambda_n = \frac{g_n^2}{\hbar \omega_n} {\chi(0)}
\end{equation}

\noindent where $\chi(0)$ is the electron-hole response function. 
Actually, the variation of atomic distances due to the excitation of the mode $n$ induces a 
modulation of electronic transfer integrals inside the system under consideration, 
here the Cu$_2$(CN)$_5$H$_4$ cluster. 
This stimulates additional charge transfer and electron-hole pair creation. 

For an insulating system with a localized or negligibly dispersing (Einstein) phonon,  
the electron-hole response function at $q_n=0$ and $\omega_n$=0 can be derived in the 
frame of the dimer-charge oscillation model \cite{Rice, McDonald}: 

\begin{equation}
\chi(0)=\sum\limits_{i,j}{}{\frac{|\langle i |\hat{o}| j \rangle|^2}{\hbar \omega_{i,j}}}
\end{equation}

\noindent where the sum is performed over all possible electronic transition from the 
occupied state $i$ to the unoccupied state $j$ which are susceptible to 
be involved with the excitation of the phonon $n$. 
Supposing that the matrix elements of relevant processes are close to the unity, 
and all others are close to zero, $\chi(0)$ simplifies to:

\begin{equation}
\chi(0) = \frac{\mathcal{N}}{\hbar \omega_{CT}}
\end{equation}
 
\noindent $\mathcal{N}$ being the number of electronic transitions with energy 
$\hbar \omega_{CT}$ which participate to the EPC. 

A cyanide CN group phonon in k-ET-Cu modulates electronic 
transfer integrals inside the anion layer as well as between the anion 
and the cation layer. 
Two lowest energy electronic excitations in the cluster are chosen as charge transfer 
excitations which are related to the modulation of intra-layer and inter-layer electronic 
transfer integrals: 
the $\pi_{\perp}$~$\rightarrow$~$\pi_{\perp}^{*}$ and 
$\pi_{\parallel}$~$\rightarrow$~$\pi_{\perp}^{*}$. 
The value of $\mathcal{N}$ is thus 2, and the expression of $\lambda_n$ becomes :

\begin{equation}
\lambda_{a} = {\frac{2 g_n^2}{(\hbar \omega_n) (\hbar \omega_{CT})}}
\end{equation}

\noindent where $\hbar \omega_{CT}$ corresponds to the minimal charge transfer excitation energy. 
It can be estimated from the calculation of the inelastic part of the spectra, in particular 
the energy of the $\pi_{\parallel}$ $\rightarrow$ $\pi_{\perp}^*$ and 
$\pi_{\perp}$ $\rightarrow$ $\pi_{\perp}^*$ excitations (structure A). 
The former is an in-plane charge transfer and the latter is possibly involved in the 
interaction with the ET-molecular layer. 
For the estimation of the $\hbar \omega_{CT}$ we took the value from the calculated spectra 
with detuning $\Omega$ = -0.4 eV, shown in Fig.~\ref{fig_9}.
The value of $\hbar \omega_{CT}$~=~5~eV corresponds to the energy loss of 
the structure A, i.e. its position relative to the elastic peak, while its  
half-width-at-half-maximum is taken as the uncertainty $\Delta\hbar \omega_{CT}$~=~0.5~eV. 
Taking this value into account, for the CN stretching mode (h$\omega_1$ = 280 meV), 
with HR parameter $S_1$ = 0.334, and $g_1$ = 230 meV, we estimate the value of the  
dimensionless EPC parameter to be $\lambda_1$ = 0.076. 

Tab.~\ref{tab2} shows the dimensionless EPC parameters of all the anion plane modes 
strongly excited by the N K edge RIXS process. 
They are individually lower than 0.08, but their total $\lambda_{anion}$ $\approx$ 0.1. 

It is noteworthy that from IR and Raman measurements \cite{McDonald} the anion plane 
modes are considered to have negligible $\lambda$ 
in the sister compound $\kappa$-(ET)$_2$Cu(SCN)$_2$. 
The same measurements reveal that the ET-molecular CC stretching mode is strongly coupled 
with $\lambda_{CC}$ = 0.17.
Similar measurements on $\beta$-(ET)$_2$I$_3$ estimate the total ET-molecular layer 
contribution to the EPC is $\lambda_{ET}$ $\approx$ 0.4 \cite{Girlando}. 
Our estimation of the EPC in $\kappa$-(ET)$_2$Cu$_2$(CN)$_3$ shows that the coupling 
of the anion modes is certainly not negligible. 
Their contribution to the total $\lambda$ is at least 20\%.

Although the anion-donor coupling is not included in the present calculation, 
it is important to realize that the nitrogen displacement induces two types of 
indirect effects on the electronic structure of the ET layer. 
First, it tilts the H-bond, inducing (small) displacements of 
the ET molecule, which modulates the inter-ET transfer integrals.
Second, it varies the H-bond distance and induces H bond polarization effect, 
which modulates inner $\sigma$ electron density of the occupied levels of the 
ET-molecule \cite{Alemany2012}.
This intra-molecular electron transfer, induces a modulation of the $\pi$-hole 
density on the core of the ET-molecules.
A similar charge modulation process was previously invoked as a consequence of the 
establishment of H bonds accompanying the charge ordering transition in several 
family of ET salts  \cite{Alemany2012, Alemany2015}. 

Our data show that the EPC of N$_C$ sites, more strongly H-bonded to the ET-ethylene 
endgroups, involves almost solely the CN stretching motion. 
Besides, the EPC of the N$_B$ sites involves all cluster CN modes depicted in Fig.~\ref{fig_4}. 
This complex dynamics, together with the orientational disorder 
of CN$_B$ cyanides, creates a substantially disordered environment of each ET molecule, 
which is both statical and dynamical. 
The observation of the near-neighbor and variable range hopping mechanisms of conductivity 
\cite{Pinteric,Culo} and the glassy behavior in dielectric spectroscopy \cite{Pinteric} 
indicates that such a disorder is certainly relevant. 
We note that charge degrees of freedom play an important role in the physics of 
k-ET-Cu, meaning that one should go beyond the simple spin-liquid description 
originally considered \cite{Shimizu}, which takes into account only magnetic 
frustration between localized spins.

\section{\label{sec:Concl}{Conclusion}}

We showed that the N K edge RIXS permits to clearly observe up to five 
lattice vibration harmonics of selected modes in $\kappa$-(BEDT-TTF)$_2$Cu$_2$(CN)$_3$ 
spin-liquid insulator. 
Tuning the incoming photon energy to the K edge of two non-equivalent N sites 
switches-on different sets of phonon modes. 
CN stretching mode is strongly excited at the edge of the ordered nitrogen 
sites which are more 
strongly connected to the BEDT-TTF molecules ($h\nu$ = 399.8 eV), 
while positionally disordered sites show multi-mode excitation ($h\nu$ = 398.4 eV). 
The difference in their resonant energy ($h\nu$) demonstrates that their environment alters 
the electronic state, while the complete change of their vibrational progression points 
to the degree of preventing the mechanical movement.
RIXS spectra are reproduced by calculations on a cluster Cu$_2$(CN)$_5$H$_4$ cut from the 
anion plane and terminated by H atoms. 
They permit to estimate that the contribution of the anion-plane modes to the 
total dimensionless electron-phonon coupling parameter of the 
$\kappa$-(BEDT-TTF)$_2$Cu$_2$(CN)$_3$ compound is about 20\%.

\section{\label{sec:ACK}{Acknowledgement}}

S.T. acknowledges the support of the Croatian Science Foundation under the project 
IP-2013-11-1011. K. M. and K. K. acknowledge the support of Japan society for the
Promotion of Science (JSPS) KAKENHI under Grant No. 25220709.


\begin{thebibliography}{99}
\bibitem{errors} Standard error of DFT based calculations of vibrations is taken as the 
uncertainty of wave numbers $\Delta \sigma_n$ (and thus energy quanta $\Delta \hbar\omega_n$) 
of each mode. 
$\Delta S_n$ ($\Delta g_n$) is derived from the error of the relative shift of 
the excited and the initial state potential. 
$\Delta\lambda_n$ includes, besides $\Delta g_n$ and $\Delta\hbar\omega_n$, 
$\Delta\hbar\omega_{CT}$ explained in the Sec.\ref{sec:EPC}.
\end{thebibliography}
\end{document}